\documentclass{iaus}
\usepackage{graphicx}

\newcommand{\MoH}{\ensuremath{\left[\mathrm{M}/\mathrm{H}\right]}}

\title[Convection in late-type giants] %% give here short title %%
{Convection and observable properties of late-type giants}

\author[A. Ku\v{c}inskas, H.-G. Ludwig, P.H. Hauschildt]   %% give here short author list %%
{A. Ku\v{c}inskas$^1$$^2$, H.-G. Ludwig$^3$ \and P.H. Hauschildt$^4$}

\affiliation
{
$^1$National Astronomical Observatory of Japan, Mitaka, Tokyo 181-8588, Japan
    \break email: arunas.kucinskas@nao.ac.jp\\[\affilskip]
$^2$Institute of Theoretical Physics and Astronomy, Go\v {s}tauto 12, Vilnius 01108, Lithuania
    \break email: ak@itpa.lt\\[\affilskip]
$^3$GEPI - CIFIST, Observatoire de Paris-Meudon, 5 place Jules Janssen, \break 92195 Meudon Cedex , France\\[\affilskip]
$^4$Hamburger Sternwarte, Gojenbergsweg 112, 21029 Hamburg, Germany
}

\pubyear{2005}
\volume{232}  %% insert here IAU Symposium No.
\pagerange{1--6}
\date{}
\jname{The Scientific Requirements for Extremely Large Telescopes}
\editors{P. Whitelock, B. Leibundgut \& M. Dennefeld, eds.}
\begin{document}

\maketitle

\begin{abstract}
We show that contrary to what is expected from 1D stationary model
atmospheres, 3D hydrodynamical modeling predicts a considerable
influence of convection on the spectral properties of late-type
giants. This is due to the fact that convection overshoots into
the formally stable outer atmospheric layers producing a notable
granulation pattern in the 3D hydrodynamical models, which has a
direct influence on the observable spectra and colors. Within the
framework of standard 1D model atmospheres the average thermal
stratification of the 3D hydro model can not be reproduced with
any reasonable choice of the mixing length parameter and
formulation of the turbulent pressure.  The differences in
individual photometric colors -- in terms of 3D versus 1D -- reach
up to $\sim0.2$\,mag, or $\Delta T_{\rm eff}\sim70$\,K. We discuss
the impact of full 3D hydrodynamical models on the interpretation
of observable properties of late-type giants, briefly mentioning
problems and challenges which need to be solved for bringing these
models to a routine use within the astronomical community in 5-10
years from now.



\end{abstract}

\firstsection % if your document starts with a section,

\section{Introduction}

Convection plays an important role in governing the interior
structure and evolution of late-type giants (i.e., stars on the
red and asymptotic giant branches, RGB/AGB). Besides of aiding the
energy transport from stellar interiors to the outer layers,
convection is important in delivering heavy elements from the
nuclear burning layers to the outer atmosphere. Since convective
mixing changes the chemical composition both in the outer
atmosphere and stellar interiors, it eventually alters also the
stellar structure because of changes in atomic and molecular
opacities. This affects observable properties of a star, lifetimes
in different evolutionary stages, and so forth. Obviously, proper
understanding of convection is of fundamental importance for
building realistic evolutionary models, which form the basis of
our understanding of individual stars and stellar populations.

Convection in current theoretical models is treated in a rather
simplistic way, typically within a framework of mixing length
theory (MLT). Inevitably, this has a number of drawbacks. For
instance, the efficiency of convection within the framework of MLT
is scaled by a-priori unknown mixing-length parameter,
$\alpha_{\rm MLT}$ (defined as the ratio of the mixing length to
the pressure scale height), usually calibrated with Solar models.
Almost certainly, $\alpha_{\rm MLT}$ needs not to be the same in
main-sequence stars, subgiants, giants and supergiants, as is
assumed in current evolutionary models (see, e.g., \cite[Freytag
\& Salaris 1999]{FS99} for more details on this issue). Not
surprisingly, there have been many attempts during the last few
decades to improve the treatment of convection in stellar models
(for example, implementing the concept of convective overshooting,
which allows for convection to penetrate beyond the classical
boundaries of a convective layer). While such efforts are
incremental steps towards a more realistic modeling of convection
in stellar interiors, a fundamental breakthrough in this area is
likely to go beyond the classical stationary 1D modeling.
Obviously, this may be possible with full 3D hydrodynamical
models, as they account for time-dependent and three-dimensional
character of convection from first principles, providing a degree
of realism in the treatment of non-stationary phenomena (and
convection in particular) that is beyond reach with classical
approach.

\begin{figure}
\centering
\includegraphics[width=6cm]{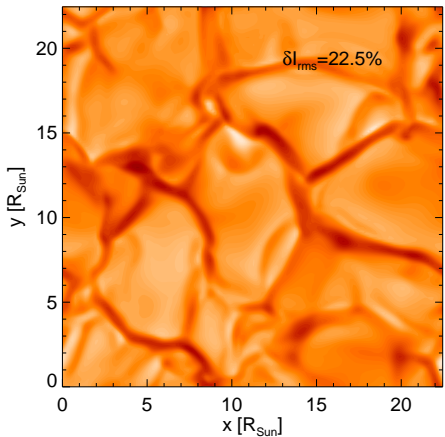}
\caption{Snapshot of the emergent white light intensity during the
temporal evolution of a hydrodynamical red giant model. Note the
spatial scales of the granulation pattern. The relative rms
intensity contrast of the granulation pattern is 22.5\,\% at this
particular instant in time.}\label{fig:3Dsurface}
\end{figure}

\section{Convection in late-type giants with 3D hydrodynamical models}

To investigate the influence of convection on the observable
properties of late-type giants we employed a
radiation-hydrodynamics code named {\tt CO$^5$BOLD} (mainly
developed by B.~Freytag and M.~Steffen; for a description of the
code see \cite[Wedemeyer \etal\ 2004]{WFSLH04}) to construct a
fully time-dependent 3D hydrodynamical model of a prototypical
late-type giant ($T_{\rm eff}\simeq 3700$\,K, $\log g=1.0$, ${\rm
\MoH}=0.0$; Fig.~\ref{fig:3Dsurface}). The final goal of this work
was to make a comparison of the thermal structures and spectral
properties of the 3D model with those predicted by classical 1D
model atmospheres. Note that 1D atmosphere model used in this
comparison was calculated using the same atmospheric parameters
and employed the same physical input data (opacities, equation of
state, description of radiative transfer) as the hydrodynamical
model.

\begin{figure}
\centering
\includegraphics[width=7.5cm]{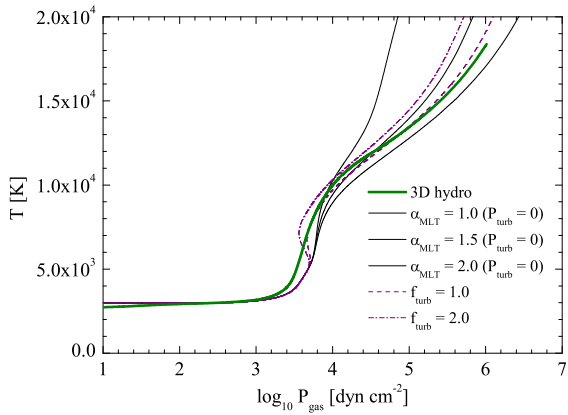}
\caption{Temperature stratification in the 3D hydrodynamical model
of a late-type giant, as a function of gas pressure, $P_{\rm
gas}$. Thick solid line is a 3D hydrodynamical model (averaged on
surfaces of equal geometrical depth), thin lines are 1D
plane-parallel models calculated using different mixing-length
parameter ($\alpha_{\rm MLT}=1.0,1.5,2.0$, from left to right; in
all cases $P_{\rm turb}=0$). Two 1D models with non-zero turbulent
pressure are given by dashed ($\alpha_{\rm MLT}=2.0$ and $f=1.0$)
and dashed-dotted ($\alpha_{\rm MLT}=2.0$ and $f=2.0$)
lines.}\label{fig:3Dstructure}
\end{figure}

Indeed, the thermal structure of the 3D model is quite different
from that obtained with the classical 1D approach
(Fig.~\ref{fig:3Dstructure}). Three 1D models shown in
Fig.~\ref{fig:3Dstructure} were calculated for different mixing
length parameters ($\alpha_{\rm MLT}=1.0,1.5,2.0$) assuming a
vanishing turbulent pressure, i.e. $P_{\rm turb}=0$. Two further
1D models were constructed including a non-zero turbulent
pressure, calculated as $P_{\rm turb}=f \rho v^{2}$, where $\rho$
and $v$ are gas density and velocity respectively, and $f$ is a
dimensionless factor (usually $f<1$). Interestingly, no reasonable
combination of $\alpha_{\rm MLT}$ and $f$ allows to construct a
classical 1D model which reproduces the structure of 3D
hydrodynamical model (Fig.~\ref{fig:3Dstructure}). In the
framework of MLT convection is confined to optically thick regions
($\log P > 3.5$), and consequently the optically thin regions are
hardly sensitive to convection. However, the geometric distance
between the upper boundary of the convective region and optical
depth unity is not large, and in the 3D hydrodynamical model
convection overshoots into the optically thin layer. This produces
a prominent granulation pattern on the stellar surface (cf.
Fig.~\ref{fig:3Dsurface}) which has a direct influence on the
emerging spectrum, and thus - the photometric colors.

\begin{figure}
\centering
\includegraphics[width=7cm]{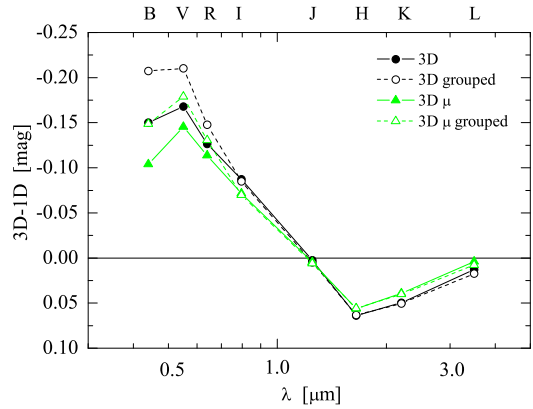}
\caption{Influence of surface granulation on the broad-band
photometric colors of red giant, as reflected by magnitude
differences in various band-passes (indicated on the top of the
panel) between the predictions of 3D hydrodynamical and classical
1D model atmospheres. The different lines depict various
approximations employed in the treatment of the 3D radiative
transfer (see \cite[Ku\v{c}inskas \etal\ 2005]{K05} for
details).}\label{fig:3Dcolors}
\end{figure}

The differences between colors predicted with 3D and 1D models are
significant, e.g., $\Delta(V-K)\sim0.2$, or $\Delta T_{\rm
eff}\sim70$\,K (Fig.~\ref{fig:3Dcolors}; see \cite[Ku\v{c}inskas
\etal\ 2005]{K05} for details on the calculation of spectra and
colors with the 3D model). Note that this difference is comparable
to a typical error margin in the individual $T_{\rm eff}$--color
relations (\cite[Ku\v{c}inskas \etal\ 2006]{K06}) which are used
to transform theoretical isochrones to color-magnitude planes.

\section{Conclusions}

Clearly, our 3D model of a prototypical late-type giant predicts
considerably different thermal structures from those inferred with
the classical 1D model atmospheres. Spectral properties of the 3D
model are rather sensitive to convection too, which results in
significant differences between photometric colors calculated with
the 3D and 1D model atmospheres. All this should be properly taken
into account both with evolutionary models and model atmospheres
of late-type giants. At the same time, there is a number of issues
that will have to be tackled to improve the 3D models, just to
name a few:

\begin{itemize}

\item
Implement a possibility of direct spectral synthesis with 3D
stellar atmosphere models, preferably in non-LTE;

\item
Improve spectral line databases, atomic and (especially!)
molecular opacities;

\item
Investigate the properties of stars in the regions of HR diagram
that were poorly covered with 3D models up to now, at different
metallicities (especially RGB/AGB stars).

\end{itemize}

While 3D hydrodynamical models are still computationally expensive
today, grids of synthetic spectra calculated in 3D approach may be
available in 5-10 years from now (see \cite[Ludwig \&
Ku\v{c}inskas 2005]{LK05} for a discussion). No doubt, this will
open new possibilities for improving models of stellar evolution
(especially those on RGB/AGB), and will alow to study a variety of
new phenomena that are beyond reach with classical 1D models.

\begin{discussion}

\discuss{Feast}{At least in case of large amplitude variables one
can circumvent the models by empirical calibration of luminosity,
age and metallicity in our Galaxy and the Local Group.}

\discuss{Ku\v{c}inskas}{Calibration of selected stellar properties
is indeed possible with observations. However, all three
quantities you mention are derived employing certain theoretical
assumptions, thus they are not entirely model-independent.}

\discuss{Feast}{What is it in your models which determines the
size of the convective cells?}

\discuss{Ku\v{c}inskas}{To first approximation the size of the
convective cells is set by (five to ten times) the pressure scale
height at the stellar surface. In earlier spectral types with more
intense radiation fields the cell size is somewhat biased towards
larger values. Sphericity related effects in late-type supergiants
are not fully clarified yet. However, there exist indications of a
bias again towards larger cells.}

\discuss{Ardeberg}{With the new synthetic spectra and colors
available, should we foresee any problems related to our current
models, simulations and conclusions?}

\discuss{Ku\v{c}inskas}{I would expect that predictions of the
current theoretical models will not change drastically. There is
no doubt, however, that certain details of the global picture will
be different. How much - this needs to be clarified with further
modeling.}

\end{discussion}

\end{document}